\pdfoutput=1
\documentclass{sigchi-ext}[]
\usepackage[T1]{fontenc}
\usepackage{textcomp}
\usepackage[scaled=.92]{helvet} 
\usepackage{graphicx} 
\usepackage{subfigure}

\usepackage{balance}  
\usepackage{booktabs} 
\usepackage{ccicons}  
\usepackage{ragged2e} 

\copyrightinfo{Permission to make digital or hard copies of part or all of this work for personal or classroom use is granted without fee provided that copies are not made or distributed for profit or commercial advantage and that copies bear this notice and the full citation on the first page. Copyrights for third-party components of this work must be honored. For all other uses, contact the Owner/Author. 
Copyright is held by the owner/author(s). \\
{\emph{CHI'17 Extended Abstracts}},  May 06-11, 2017, Denver, CO, USA. \\
ACM 978-1-4503-4656-6/17/05.\\
http://dx.doi.org/10.1145/3027063.3053227
}

\title{Automated Assistants to Identify and Prompt Action on Visual News Bias}

\numberofauthors{7}
\author{%
    \alignauthor{%
        \textbf{Vishwajeet Narwal}\\
        \affaddr{West Virginia University (WVU)} \\
        \affaddr{vnarwal95@gmail.com} }
    \alignauthor{%
        \textbf{John O'Donovan}\\
        \affaddr{University of California, Santa Barbara (UCSB)}\\
        \email{jod@cs.ucsb.edu} } \vfil 
    \alignauthor{%
        \textbf{Mohamed Hashim Salih}\\
        \affaddr{Universidad Nacional Autonoma de Mexico (UNAM)}\\
        \email{mhssulimanm@gmail.com} }
    \alignauthor{%
        \textbf{Tobias H{\"o}llerer}\\
        \affaddr{University of California, Santa Barbara (UCSB)}\\
        \email{holl@cs.ucsb.edu} } \vfil 
    \alignauthor{%
        \textbf{Jose Angel Lopez}\\
        \affaddr{Universidad Nacional Autonoma de Mexico (UNAM)}\\
        \email{joseangel.robotics@gmail.com} }
    \alignauthor{%
        \textbf{Saiph Savage}\\
        \affaddr{West Virginia University (WVU)}\\
        \email{saiph.savage@mail.wvu.edu} } \vfil
    \alignauthor{%
        \textbf{Angel Ortega}\\
        \affaddr{University of California, Santa Barbara (UCSB)}\\
        \email{angelortega@cs.ucsb.edu} }
    }

\def\plaintitle{An Automated Assistant to Identify and Prompt Action on Visual Bias in News} \def\plainauthor{Vishwajeet Narwal, Mohamed Hashim Salih, Jose Angel Lopez, Angel Ortega, John O'Donovan, Tobias Hollerer, Saiph Savage}
\def \plainkeywords{\small News Bias; Fake News; Civic Computing}

\hypersetup{%
  pdftitle={\plaintitle}, pdfauthor={\plainauthor},
  pdfkeywords={\plainkeywords}, }

\begin{document}

\maketitle

\RaggedRight{} 

\begin{abstract}
Bias is a common problem in today's media, appearing frequently in text and in visual imagery. Users on social media websites such as Twitter need better methods for identifying bias. Additionally, activists --those who are motivated to effect change related to some topic, need better methods to identify and counteract bias that is contrary to their mission. With both of these use cases in mind, in this paper we propose a novel tool called UnbiasedCrowd that supports identification of, and action on bias in visual news media. In particular, it addresses the following key challenges (1) identification of bias; (2) aggregation and presentation of evidence to users; (3) enabling activists to inform the public of bias and take action by engaging people in conversation with bots. We describe a preliminary study on the Twitter platform that explores the impressions that activists had of our tool, and how people reacted and engaged with online bots that exposed visual bias. We conclude by discussing design and implication of our findings for creating future systems to identify and counteract the effects of news bias.
 
\end{abstract}

\keywords{\plainkeywords}

\category{\small K.4.2}{\small Computers and Society}{\small Social Issues}

\section{Introduction}

News bias refers to the bias that journalists have in how they report and cover certain events and stories \cite{kang2015full,parry2010visual}. Visual bias is a type of news bias, where images are used to accentuate or play down certain aspects of a story. For instance, when covering a protest, the media might strategically select photos that show violence to promote a negative perspective about the event, even when the protest was largely peaceful \cite{arpan2006news,corrigall2011picturing}. Previous research has examined a broad and diverse set of world events in which bias plays significant roles. These include racial stereotyping \cite{ waters2016threats}, coverage of war news \cite{parry2010visual}, protests \cite{corrigall2011picturing}, natural disasters \cite{borah2009comparing} and elections \cite{coleman2006network}.  More recently, we have seen many reports of bias on both sides of the ``Brexit'' debate in the United Kingdom, e.g. \cite{brexit}, and on both sides of the US presidential election, e.g. \cite{uselect}.  These examples highlight a need for better methods to identify and mitigate news bias. 
\begin{marginfigure}[4pc]
  \begin{minipage}{\marginparwidth}
    \centering
    \includegraphics[width=0.9\marginparwidth]{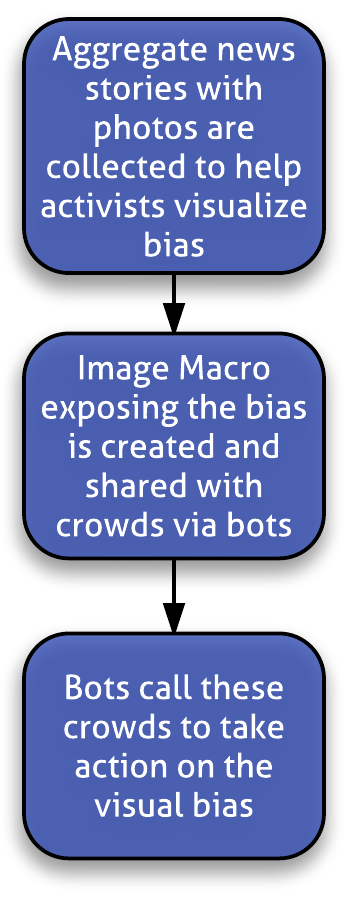}
    
    \caption{Overview of how UnbiasedCrowd functions.}~\label{fig:overview}
  \end{minipage}
\end{marginfigure}

Although considerable research has been devoted to understanding how visual bias is present in real world news, less attention has been paid to developing systems that can automatically detect visual bias and enable people to mitigate its effect. Until recently have we seen the emergence of tools to detect possible biases. PageOneX \cite{costanza2016pageonex} helps uncover placement bias. CrowdVoice and Ushahidi \cite{okolloh2009ushahidi} help activists to compare different sources of news. Other systems extract news stories from different sources to try and help people to consider a story from different viewpoints and  formulate a more balanced perspective \cite{park2009newscube, chhabra2012cubethat,iacobelli2010tell}. While these tools do offer a relative ease to detect possible biases in news, they are not envisioned to take action after the bias has been detected. Given that the goals of activists is to drive change  \cite{arpan2006news,corrigall2011picturing}, we believe it is especially important to develop tools that can help not only normal users but also activists to incite actions against news biases.

In this late-breaking work, we introduce \emph{UnbiasedCrowd}, a system that helps typical users to identify bias, and also helps activists to easily detect visual bias, expose the bias to people, and prompt collective action on the bias. Figure \ref{fig:overview} presents an overview of our system.  

\begin{figure}
  \includegraphics[width=0.90\columnwidth]{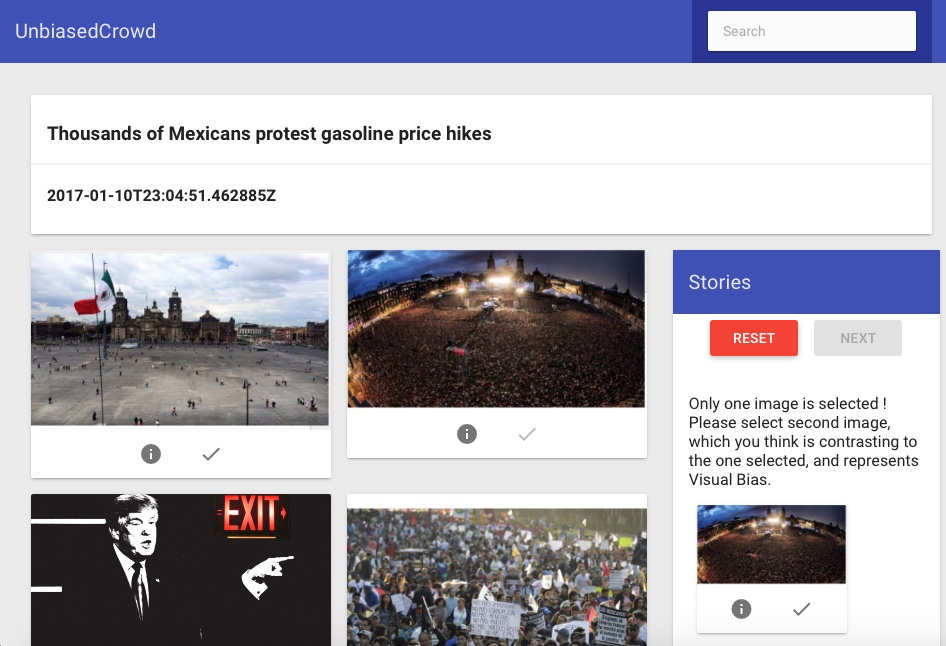}
  \caption{UnbiasedCrowd's Interface that shows all aggregated images of a story to help activists detect the bias and select images to create the image macro that will be shared on Twitter.}~\label{fig:interface2}
 
\end{figure}
Given its action aspect, we view UnbiasedCrowd as an automated assistant to activists because the system helps activists to drive people to do something about visual biases. We ran a pilot study to understand how people react to our system that exposes them to visual biases and nudges them to take action. We found that individuals decided to engage differently with the system. Some individuals acknowledged the presence of bias in news article and took action against it, while others advocated the use of certain photos by media. 
We discuss design implications of our results, and engage in a discussion of systems that expose and help drive action in citizens towards news biases. 

\section{UnbiasedCrowd System}

Our system consists of 3 modules: 1) bias detection; 2) bias exposure; and 3) call to action (see figure \ref{fig:overview}).

\emph{\bf 1. Bias Detection Module.} This module collects all of the images of the news story that the user specifies, clusters them, and displays them back to help the user detect possible biases. In our design, we consider that the user, in general, will be an activist who is interested in detecting visual biases and mobilizing people to take action on the bias. 
\begin{marginfigure}[-4pc]
  \begin{minipage}{\marginparwidth}
    \centering
    \includegraphics[width=1.0\marginparwidth]{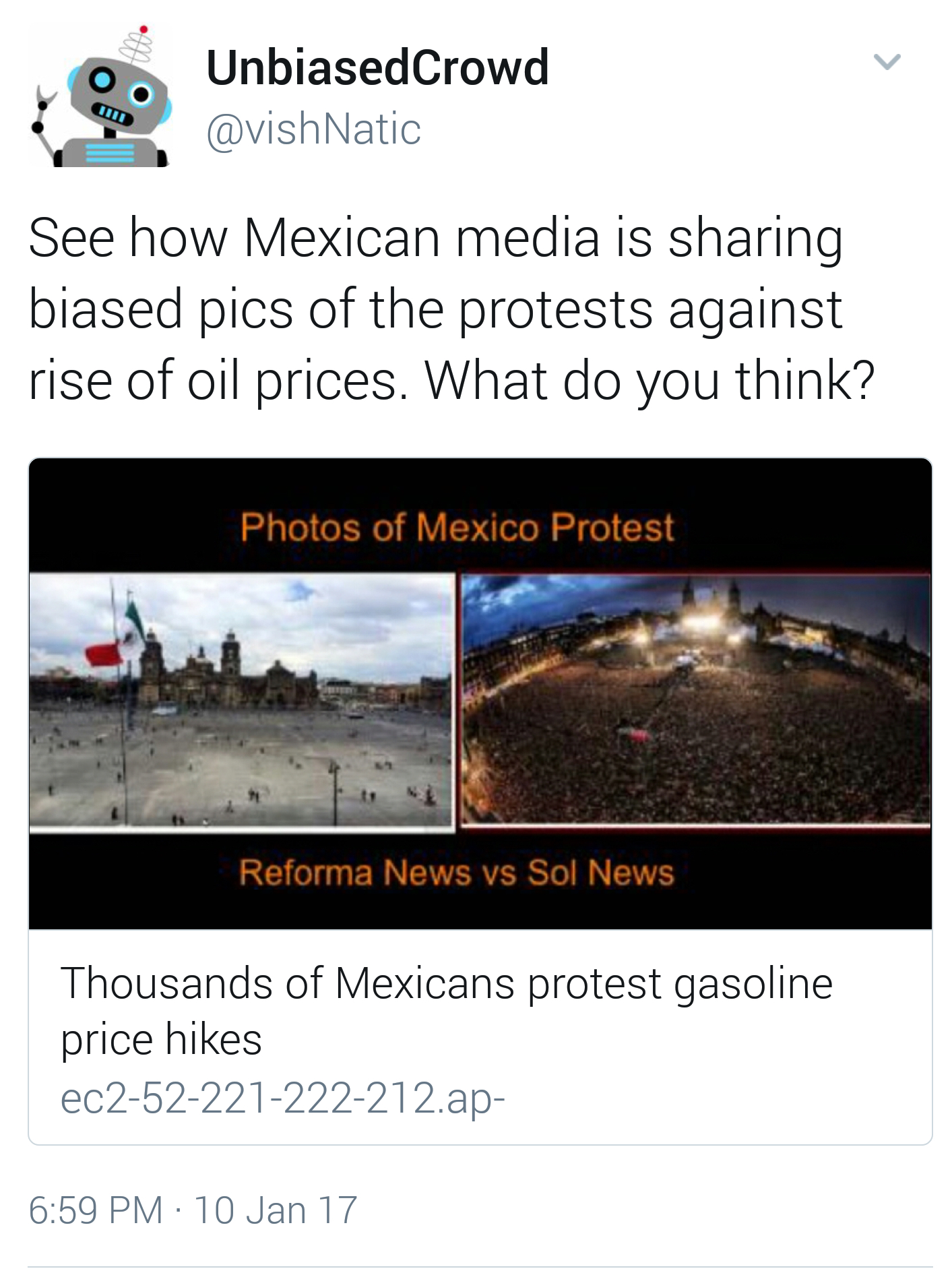}
    \caption{Bot shares the image macro on Twitter to inform people about the possible bias}~\label{fig:tweet}
  \end{minipage}
\end{marginfigure}
\begin{marginfigure}[1pc]
  \begin{minipage}{\marginparwidth}
    \centering
    \includegraphics[width=1.0\marginparwidth]{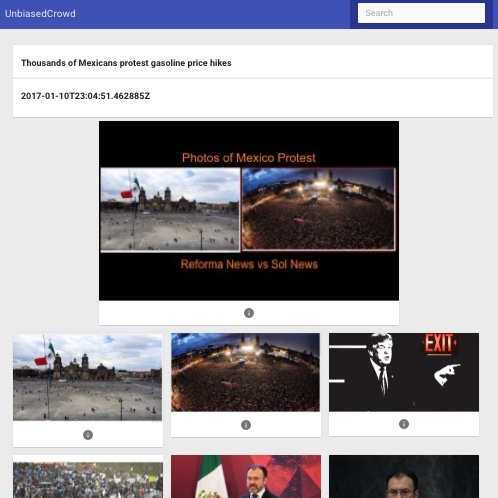}
    \caption{Upon click of the image macro, people are re-directed to a site where they can explore for themselves all the media that has been generated for a particular story, and thus make the decision themselves whether there is indeed a bias or not. }~\label{fig:interface4}
  \end{minipage}
\end{marginfigure}
Our system depends on Google's relevance algorithm to collect articles from different sources belonging to one same story, and then extract the photos from these news links directly. Once the images are collected we use standard computer vision techniques to cluster similar images together (in particular, we calculate each image's Fisher vector and do K-means clustering). Activists can start the investigation by looking over the image clusters and seeing if there are any discrepancies in how the media is covering an event (see Figure \ref{fig:interface2}).
Having clusters of images to compare helps activists to more easily detect visual bias as visual bias can be understood by comparing many images together instead of only one image. 

\emph{\bf 2. Bias Exposure Module.} 
After activists have detected a visual bias in a particular news story, our system helps them to create an image macro/collage to share with crowds to expose them to the media bias. 
\begin{figure}
\begin{center}
  \includegraphics[width=0.8\columnwidth]{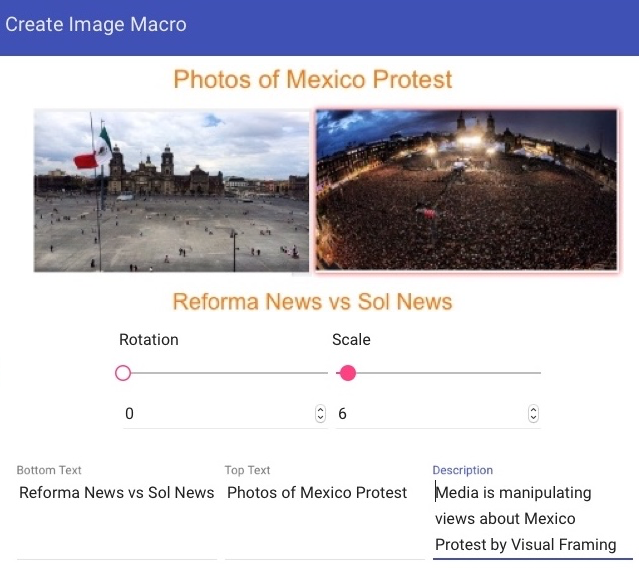}
  \caption{UnbiasedCrowd's Interface for creating an image macro using selected set of images. }~\label{fig:interface3}
  \end{center}
\end{figure}
\enlargethispage{\baselineskip}

Figure \ref{fig:interface3} shows our system's interface to help activists create this image macro or collage. After the collage has been created, our system then uses online bots to share the image macro on social media (see Fig. \ref{fig:tweet}). In particular, the image is shared as a Twitter Summary Card\footnote{\small \url{https://dev.twitter.com/cards/types/summary-large-image}}, which on clicking, redirects to a page on our system's server  where  all the aggregated photos of the story are presented (see Fig. \ref{fig:interface4}). This  enables people to investigate the case by themselves (people have access to the collage that the activist created, and also to all the images that our system collected). Consequently, people are empowered to detect if the activists, themselves, are trying to bias them. 

\begin{marginfigure}[0pc]
  \begin{minipage}{\marginparwidth}
    \centering
    \includegraphics[width=1.0\marginparwidth]{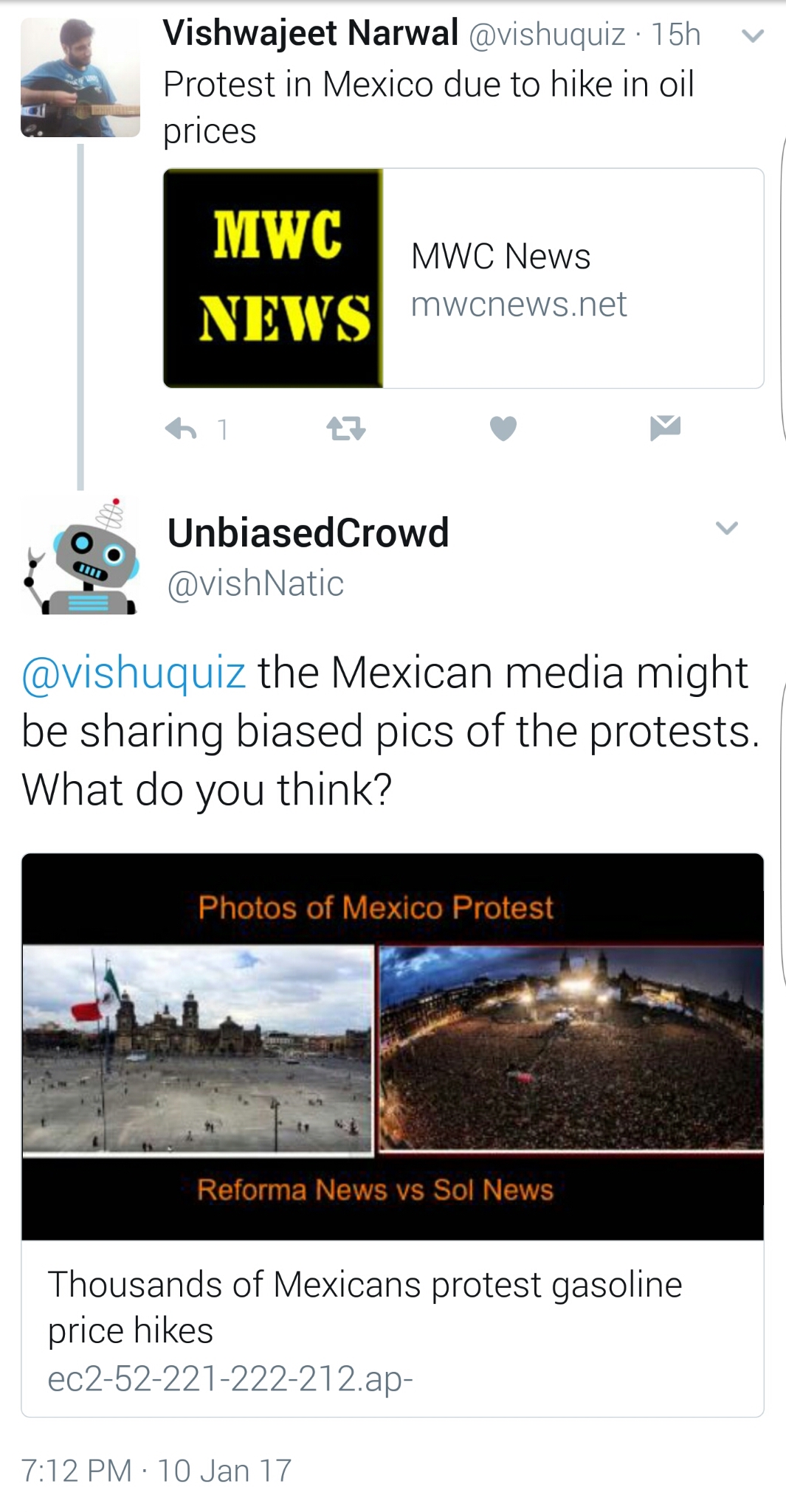}
    
    \caption{Bot interacting with a user who has shared an article susceptible to visual bias}~\label{fig:tweet_r}
  \end{minipage}
\end{marginfigure}

Figure \ref{fig:tweet_r} showcases how the bots of our system engage with social media users to expose them to the bias. In our current implementation, our system searches for people who are currently tweeting about the news story (based on hashtags or news article they are sharing online.) 
We considered these individuals could be the most interested in being aware of the bias. Our bots then simply call people out and present to them the constructed image macro, asking them if they think there is a possible bias.

\emph{\bf 3. Call to Action Module} After people have been exposed to the bias, the online bots of our system engage in conversation with people to try and drive them to take some type of action on the bias.  In particular, the bots start to communicate with the people who were originally targeted (i.e., the individuals to whom the bots exposed the visual bias) by replying to their tweets and engaging them in conversation. The conversation focuses on asking people what actions can be done given the possible news media bias and encouraging people to take action. Note that the taking action part is supervised by activists who at this point can jump in to lead people to action themselves. 

\section{Preliminary Study}

We ran (1) interviews and a (2) pilot study to understand how people reacted and engaged with our system.  

\subsection{Method} 
We recruited a real world activist group to use our system. We conducted interviews with the leaders and members of the group to gain insights on their perspectives and impressions on tools such as UnbiasedCrowd. We also conducted a pilot study to inspect how people engaged with the online bots of our system.\\
\emph{\bf Interview Study:} In our interview study, we asked activists to use UnbiasedCrowd interface for 5-15  minutes and then asked them to discuss their thoughts on the system. Participants voiced their thoughts of what they thought about our system,  how and why they would use our tool or not, and how they generally used social media. We used qualitative coding to analyze interview responses. This allowed us to establish categories of the  perceptions that activists had about our system. 

\emph{\bf Pilot Study:} We probed our system within the use case of having activist groups who wanted to expose the media bias to the general population. Activists used our system to uncover visual bias in a topic of their interest. They then used the online bots of our system to expose people to the bias, and initiate a discussion with people on the actions that should be taken. 
\enlargethispage{\baselineskip}

\subsection{Participants}
We had the following participants in our study:\\
\emph{\bf Activists:} In particular, we worked with  activist groups from Mexico\footnote{\small \url{https://www.facebook.com/AccionRevolucionariaMX/}} who fought to have a better energy reform in their country.  These activists felt that the Mexican government and news media were biased in how they portrayed the protests on the energy reforms. The activists thought that the government wanted people to think that nobody was taking action or doing anything about the energy reform to demotivate citizens to participate in opposition movements.

\emph{\bf General Population:} For our study, our bots targeted people on Twitter to expose and discuss with them the possible visual news bias in Mexico's energy reform protests. Our bots particularly targeted people who were already discussing on social media Mexico's energy reform. We targeted these individuals given that related work has shown they might be more likely to respond and participate in  collective action \cite{savage2016botivist}. All people that our bots targeted spoke Spanish and the messages of our bots were also in Spanish. Activists helped tailor the messages of the bots. Bots looked for people who were mentioning hashtags related to the Mexican energy reform such as \emph{``\#Gasolinazo'', ``\#ReformaEnergetica'', ``\#PEMEX''}. 

To engage the general public, we put into operation one of the bots of UnbiasedCrowd on Twitter, which made tweets where it presented the image macro that the activists had generated and asked people whether they thought there was a bias in how the news media was reporting the energy reform in Mexico.  If a person replied or mentioned the bot, the bots replied with canned questions on  what actions should be done about the possible visual bias. Note that we followed the design of \cite{savage2016botivist} to call and initiate action with people. We took rigorous steps to ensure ethical practices. People that our bots targeted were informed about the study after the fact and could opt out of participating (in which case we did not use their data.) All of the individuals that our bots targeted allowed us to use their data.

\section{Results} 

\subsection{Qualitative Results: Interviews}
We had a total of 3 individuals participate in our interview study. All interviewees expressed they considered our interface was useful to expose people about the bias and start to discuss best ways to take action. The following categories covered participants' perspectives of our system. 

{\bf Transparency:} All participants liked the fact that our system allowed people to view and inspect the collection of photos that were collected from the news stories. They felt that having transparency about where the data came from, helped make their cause and arguments seem more truthful. Some activists also expressed that our system's site should state that it worked independently of the activists' cause, again to help them seem more authentic and real. 

{\bf Education:} Our interviewees expressed interest in expanding our tool to be more educational. They wanted to use our system to help people learn in a step by step process about visual bias, especially to empower people to detect it on their own and take action without any prompts. We also believe this is an interesting and important direction. In the future, we plan on exploring how our tool could be used to create micro-learning opportunities about detecting and taking action on news bias.  

{\bf Context:} Our interviewees also expressed that they wanted to have a way to provide context to the clusters of images that our system generated. For instance, interviewees  wanted to explain to the general public that all the newspapers that support the government were showcasing empty streets to potentially make people feel that there was no issue with the energy reform. We have considered that allowing activists to provide annotations to the clusters of images could be informative for others. However, allowing annotations might not allow people to make their own decisions on what certain images might actually mean, and this could affect how people interpret the visual bias. This could even lead to people being potentially manipulated by the activists themselves. For future work, we would like to explore the consequences of allowing  annotations in the clusters of images that our system generates. We are especially interested in studying how it affects how people understand the bias. We would also like to explore allowing annotations from people with different perspectives.

\subsection{Quantitative Results: Pilot Study}
Our bots targeted 30 people on social media and received a total of 53 responses from different individuals. Most people replied at least twice to the bots. We manually inspected the replies that people gave to the bots. People's first response was usually about their opinion on the visual bias that the bots were presenting to them; and the second reply was about the actions or next steps they thought should be done. 
We found that people, in general, took on two different types of actions. Some individuals appeared to take action as ``evangelists'' and wanted their friends and contacts to also see how they might have been biased by the media. Other individuals, appeared to take a defense stand, and actively tried to justify why it made sense for the news media to use certain photos for a story.

\section{Future Work}

It is not possible to find and recruit activist for each and every possibly biased news story, which implies the importance of educating the general public to detect Media Bias on their own. We would like to explore the possibility of using work of activists on our platform to train general users and/or Crowdworkers to be able to detect bias. We will also conduct a follow-up study to analyze how our system affects how people understand and engage with visual news bias. We will use the results to improve the design of our system and to inform the design of future social systems that organize citizens for collective action.

{\bf Acknowledgements.} This work was partially funded by a J. Wayne and Kathy Richards Faculty Fellowship. 
\balance{} 
\small
\bibliographystyle{SIGCHI-Reference-Format}
\small
\bibliography{chi}


\begin{thebibliography}{00}


\ifx \showCODEN    \undefined \def \showCODEN     #1{\unskip}     \fi
\ifx \showDOI      \undefined \def \showDOI       #1{{\tt DOI:}\penalty0{#1}\ }
  \fi
\ifx \showISBNx    \undefined \def \showISBNx     #1{\unskip}     \fi
\ifx \showISBNxiii \undefined \def \showISBNxiii  #1{\unskip}     \fi
\ifx \showISSN     \undefined \def \showISSN      #1{\unskip}     \fi
\ifx \showLCCN     \undefined \def \showLCCN      #1{\unskip}     \fi
\ifx \shownote     \undefined \def \shownote      #1{#1}          \fi
\ifx \showarticletitle \undefined \def \showarticletitle #1{#1}   \fi
\ifx \showURL      \undefined \def \showURL       #1{#1}          \fi

\bibitem{arpan2006news}
{Laura~M Arpan}, {Kaysee Baker}, {Youngwon Lee}, {Taejin Jung}, {Lori Lorusso},
  {and} {Jason Smith}. 2006.
\newblock \showarticletitle{News coverage of social protests and the effects of
  photographs and prior attitudes}.
\newblock {\em Mass Communication \& Society\/} {9}, 1 (2006), 1--20.
\newblock


\bibitem{borah2009comparing}
{Porismita Borah}. 2009.
\newblock \showarticletitle{Comparing visual framing in newspapers: Hurricane
  Katrina versus tsunami}.
\newblock {\em Newspaper Research Journal\/} {30}, 1 (2009), 50.
\newblock


\bibitem{chhabra2012cubethat}
{Sidharth Chhabra} {and} {Paul Resnick}. 2012.
\newblock \showarticletitle{Cubethat: news article recommender}. In {\em
  Proceedings of the sixth ACM conference on Recommender systems}. ACM,
  295--296.
\newblock


\bibitem{coleman2006network}
{Renita Coleman} {and} {Stephen Banning}. 2006.
\newblock \showarticletitle{Network TV news' affective framing of the
  presidential candidates: Evidence for a second-level agenda-setting effect
  through visual framing}.
\newblock {\em Journalism \& Mass Communication Quarterly\/} {83}, 2 (2006),
  313--328.
\newblock


\bibitem{corrigall2011picturing}
{Catherine Corrigall-Brown} {and} {Rima Wilkes}. 2011.
\newblock \showarticletitle{Picturing protest: The visual framing of collective
  action by first nations in Canada}.
\newblock {\em American Behavioral Scientist\/} (2011), 0002764211419357.
\newblock


\bibitem{costanza2016pageonex}
{Sasha Costanza-Chock} {and} {Pablo Rey-Maz{\'o}n}. 2016.
\newblock \showarticletitle{PageOneX: New Approaches to Newspaper Front Page
  Analysis}.
\newblock {\em International Journal of Communication\/}  {10} (2016), 28.
\newblock


\bibitem{iacobelli2010tell}
{Francisco Iacobelli}, {Larry Birnbaum}, {and} {Kristian~J Hammond}. 2010.
\newblock \showarticletitle{Tell me more, not just more of the same}. In {\em
  Proceedings of the 15th international conference on Intelligent user
  interfaces}. ACM, 81--90.
\newblock


\bibitem{kang2015full}
{Byungkyu Kang}, {Tobias H{\"o}llerer}, {and} {John O'Donovan}. 2015.
\newblock \showarticletitle{The Full Story: Automatic detection of unique news
  content in Microblogs}. In {\em Advances in Social Networks Analysis and
  Mining (ASONAM), 2015 IEEE/ACM International Conference on}. IEEE,
  1192--1199.
\newblock


\bibitem{brexit}
{Jane Martinson}. 2016.
\newblock BBC hits back at Daily Mail accusation of 'Brexit bias'.
\newblock   (October 2016).
\newblock
\showURL{%
\url{https://www.theguardian.com/media/2016/oct/13/bbc-hits-back-against-daily-mail-accusation-of-brexit-bias}}
\newblock
\shownote{[Accessed: 05-January-2017].}


\bibitem{okolloh2009ushahidi}
{Ory Okolloh}. 2009.
\newblock \showarticletitle{Ushahidi, or 'testimony': Web 2.0 tools for
  crowdsourcing crisis information}.
\newblock {\em Participatory learning and action\/} {59}, 1 (2009), 65--70.
\newblock


\bibitem{park2009newscube}
{Souneil Park}, {Seungwoo Kang}, {Sangyoung Chung}, {and} {Junehwa Song}. 2009.
\newblock \showarticletitle{NewsCube: delivering multiple aspects of news to
  mitigate media bias}. In {\em Proceedings of the SIGCHI Conference on Human
  Factors in Computing Systems}. ACM, 443--452.
\newblock


\bibitem{parry2010visual}
{Katy Parry}. 2010.
\newblock \showarticletitle{A visual framing analysis of British press
  photography during the 2006 Israel-Lebanon conflict}.
\newblock {\em Media, War \& Conflict\/} {3}, 1 (2010), 67--85.
\newblock


\bibitem{savage2016botivist}
{Saiph Savage}, {Andres Monroy-Hernandez}, {and} {Tobias H{\"o}llerer}. 2016.
\newblock \showarticletitle{Botivist: Calling Volunteers to Action using Online
  Bots}. In {\em Proceedings of the 19th ACM Conference on Computer-Supported
  Cooperative Work \& Social Computing}. ACM, 813--822.
\newblock


\bibitem{uselect}
{John Sides}. 2016.
\newblock Is the media biased toward Clinton or Trump? Here is some actual hard
  data.
\newblock   (September 2016).
\newblock
\showURL{%
\url{http://wpo.st/IU2R2}}
\newblock
\shownote{[Accessed: 05-January-2017].}


\bibitem{waters2016threats}
{Grace Waters}. 2014.
\newblock {\em Threats, Parasites and Others: The Visual Framing of Roma
  Migrants in the British Press}.
\newblock Master's\ thesis. London School of Economics and Political Science.
\newblock


\end{thebibliography}


\end{document}